\documentclass[a4paper,fleqn,sort&compress,numbers]{cas-sc}
\usepackage{natbib}
\usepackage{subfiles} 
\usepackage{siunitx}
\usepackage{amsmath}
\usepackage{makecell}
\usepackage{placeins}
\usepackage{booktabs}
\usepackage{multicol}
\usepackage{graphicx}
\usepackage{hhline}
\def\tsc#1{\csdef{#1}{\textsc{\lowercase{#1}}\xspace}}
\tsc{WGM}
\usepackage{float}

\usepackage{lineno}
\usepackage{setspace}

\newcommand{\Nuc}[2]{\ensuremath{^{#2}\mbox{#1}}}
\newcommand{\ar}{\Nuc{Ar}{39}}

\newcommand{\ledwavelength}{\SI{285}{nm}}
\newcommand{\leakage}{$[\;<1.2\times10^{-5}\;]_\mathrm{90\%~C.L.}$}
\newcommand{\yield}{46.4(2.9)~\%}

\newcommand{\fp}{\ensuremath{F_{\text{prompt}}}}

\begin{document}
\let\WriteBookmarks\relax
\def\floatpagepagefraction{1}
\def\textpagefraction{.001}
\shorttitle{Development and characterization of a slow wavelength shifting coating for background rejection in liquid argon detectors}
\shortauthors{D. Gallacher et al.}

\title [mode = title]{Development and characterization of a slow wavelength shifting coating for background rejection in liquid argon detectors}

\author[1]{D.~Gallacher}
[
type=editor,
orcid=0000-0002-9395-0560,
]
\ead{dgallacher@snolab.ca}
\cormark[1]
\fnmark[1]
\author[4]{A.~Leonhardt}
\author[2]{H.~Benmansour}
\author[2]{E.~Ellingwood}
\author[2]{Q.~Hars}
\author[3,1,6]{M.~Ku\'zniak}
\author[1]{J.~Anstey}
\author[5]{B.~Bondzior}
\author[1]{M.\,G.~Boulay}
\author[1]{B.~Cai}
\fnmark[3]
\author[5]{P.\,J.~Dere\'n}
\author[2]{P.\,C.\,F.~Di~Stefano}
\author[1]{S.~Garg}
\author[1]{J.~Mason}
\author[4]{T.\,R.~Pollmann}
\fnmark[2]
\author[2,6]{P.~Skensved}
\author[1]{V.~Strickland}
\author[2,6]{M.~Stringer}

\address[1]{Department of Physics, Carleton University, Ottawa, K1S 5B6, ON, Canada}
\address[2]{Department of Physics, Engineering Physics \& Astronomy, Queen’s University, Kingston, ON, K7L 3N6, Canada}
\address[3]{AstroCeNT, Nicolaus Copernicus Astronomical Center, Polish Academy of Sciences, Rektorska 4, 00-614 Warsaw, Poland}
\address[4]{Department of Physics, Technische Universit\"at M\"unchen, 80333 Munich, Germany}
\address[5]{Institute of Low Temperature and Structure Research, Polish Academy of Sciences, Ok\'olna 2, 50-422 Wroc\l aw, Poland}
\address[6]{Arthur B. McDonald Canadian Astroparticle Physics Research Institute, Queen’s University, Kingston ON K7L 3N6, Canada}

\cortext[cor1]{Corresponding author}
\fntext[fn1]{Currently at McGill University, Montréal, Quebec H3A 2T8, Canada}
\fntext[fn2]{Currently at Nikhef and the University of Amsterdam, Science Park, 1098XG Amsterdam, Netherlands}
\fntext[fn3]{Currently at the Office of Research Services, Queen’s University, Kingston, Ontario K7L 3N6, Canada}

\begin{abstract}
We describe a technique, applicable to  liquid-argon-based dark matter detectors, allowing for discrimination of alpha-decays in detector regions with incomplete light collection from nuclear-recoil-like events.

Nuclear recoils and alpha events preferentially excite the liquid argon (LAr) singlet state, which has a decay time of $\sim$\SI{6}{\nano\second}. The wavelength-shifter TPB, which is typically applied to the inside of the active detector volume to make the LAr scintillation photons visible, has a short re-emission time that preserves the LAr scintillation timing.
We describe the production method and characterization of a wavelength-shifting polymeric film - pyrene-doped polystyrene - which we developed for the DEAP-3600 detector. At liquid argon temperature, the film's re-emission timing is dominated by a modified exponential decay with a time constant of \SI{279\pm14}{\nano\second}, and has a wavelength-shifting efficiency of \yield{} relative to TPB, measured at room temperature. By coating the detector neck (a region outside the active volume where the scintillation light collection efficiency is low) with this film, the visible energy and the scintillation pulse shape of alpha events in the neck region are modified, and we predict that through pulse shape discrimination, the coating will afford a suppression factor of $\mathcal{O}$($10^{5}$) against these events.

\end{abstract}

\begin{graphicalabstract}
\includegraphics[width=0.95\linewidth]{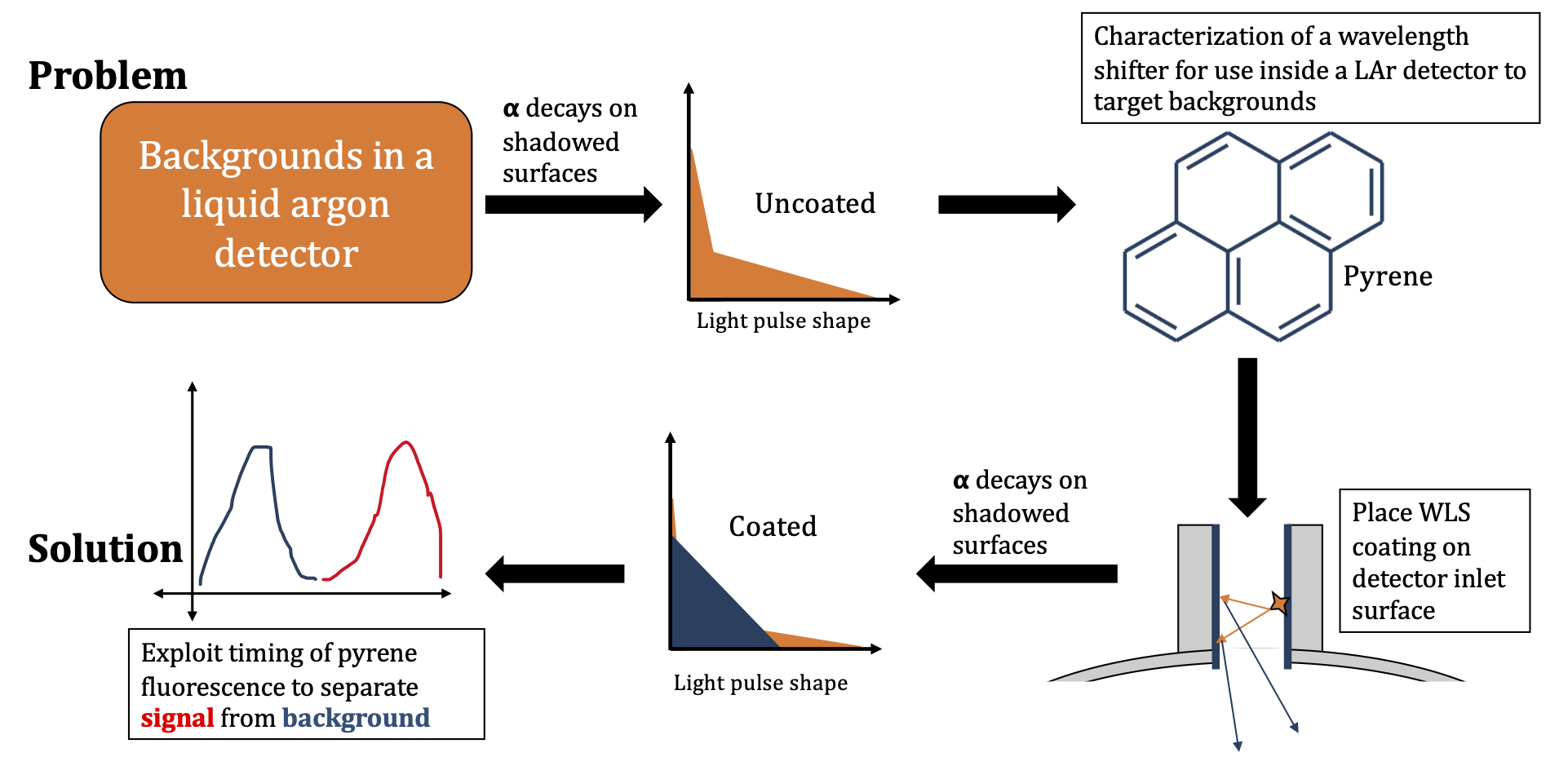}
\end{graphicalabstract}

\begin{highlights}
\item Wavelength shifter characterization
\item Background rejection techniques for liquid argon detectors
\item Cryogenic detector technologies
\end{highlights}

\begin{keywords}
Wavelength shifters
\sep
Liquid Argon
\sep
Dark Matter
\sep
TPB
\sep
Pyrene
\sep
Background Rejection
\end{keywords}

\maketitle

\section{Introduction}
\label{chap:intro}
Liquid argon (LAr) is used as a scintillator for the detection of particles in many applications. These include fundamental physics experiments for direct dark matter detection (DEAP~\cite{Amaudruz:2019fx}, DarkSide~\cite{darkside50_232d}, MiniCLEAN~\cite{Hime:2011tt}) and neutrino detection (DUNE~\cite{Abi_2020}, MicroBooNE~\cite{Acciarri_2017}, LArIAT~\cite{Acciarri_2020}). LAr scintillation light has a wavelength peaked in the vacuum ultraviolet (VUV) at $\sim$ 128 nm, below the peak efficiencies of available photo-sensors. Wavelength shifters (WLS) are materials that efficiently absorb VUV light and then re-emit photons at a longer wavelength. This emission is typically isotropic and in the blue-visible region, where commercially available photo-sensors are most sensitive. Previous studies \cite{TPBpaper,Veloce:2015slj,stanford2018surface,boulay2020} have shown that in addition to wavelength shifting, WLS coatings with scintillation decay-times sufficiently different from the LAr excimer decay-time can be used to effectively reject surface-$\alpha$ backgrounds in LAr detectors. 
In this work, we demonstrate how the wavelength shifting process itself can be used to reject alpha backgrounds originating from regions in the detector that have incomplete light collection. We describe here the development and characterization of a polymeric, WLS-doped thin-film for the rejection of alpha backgrounds from areas with limited light collection efficiency that is suitable for deployment in a liquid argon environment.

\subsection{Wavelength shifters}
 The most common WLS for LAr applications is the organic wavelength shifter 1,1,4,4-tetraphenyl-1,3-butadiene (TPB), chosen because of its high conversion efficiency ($\sim$100\% \cite{Davies:1996vd}), fast re-emission time ($\mathcal{O}$(ns) \cite{FLOURNOY1994349}), and long-term stability \cite{Burton:1973up,ArDMCollaboration:2009eu} (although emanation/solubility in LAr on the  $\mathcal{O}$(10~ppb) level was reported~\cite{Asaadi:2018vz}). TPB is an example of an organic WLS; other examples include pyrene~\cite{captalk}, p-Terphenyl~\cite{MCKINSEY1997351}, and polyethylene naphthalate~\cite{Kuzniak:2019dy}, for a recent review see~\cite{kuzniak2021wavelength}.
 
 When an incident UV photon interacts with the WLS molecule, the WLS molecule is excited from the singlet ground state, $S_{0}$, into a vibrational level of a higher-energy state $S_{1}$. The excited state can then decay down to a lower-lying state through internal conversion ($\mathcal{O}(ns)$), vibrational relaxation ($\mathcal{O}(ps)$ ), or undergo a transition to a triplet state $T_{1}$ ($\mathcal{O}(ns)$), the latter is suppressed due to its classically spin-forbidden nature. Transitions from $S_{1} \rightarrow S_{0}$ occur on $\mathcal{O}(ns)$ timescales and emit fluorescence, transitions from $T_{1} \rightarrow S_{0}$ occur on longer timescales ($\mathcal{O}(\mu s-ms)$), and produce phosphorescence. Energy loss due to internal conversion and vibrational transitions results in a wavelength shift from high energy UV light to lower energy fluorescence and phosphorescence light.
 At room temperature, the molecule may be in a vibrationally excited state, which leads to overlap between the absorption and emission spectra.

WLS are broadly characterized by the following properties:
\begin{itemize}
    \item Emission spectrum and peak wavelength(s), $\lambda_{E,max}$
    \item Photo-luminescent quantum yield (PLQY), characterized by the ratio of emitted photons/absorbed photons
    \item Re-emission lifetime(s), $\tau$
\end{itemize}

For application in LAr detectors,  the WLS must exhibit cryogenic and long-term stability for continuous operation at \SI{87}{\kelvin}. Most of the above characteristics will exhibit some dependence on temperature. At lower temperatures, vibrational modes are "frozen-in"  resulting in a smaller overlap between absorption and emission spectra, and PLQY tends to increase as thermal quenching of excited states is reduced.

\par
While TPB is the WLS material of choice for conversion of the primary scintillation light in LAr detectors, other WLS materials show strong potential for specific use within LAr applications. Pyrene is one such WLS material, and has been studied extensively in the past for different applications~\cite{Birks}. The crystal structure of pyrene is dimeric in form, and the elementary unit of the lattice is a pair of pyrene molecules. The fluorescence spectrum of pyrene has attributes of monomer ($\lambda^m_{E,max} \sim$\SI{375}{nm}) and excimer ($\lambda^e_{E,max}\sim$\SI{450}{nm}) emission, while the absorption spectrum is characteristic of monomer absorption~\cite{Birks}. Monomers that are initially excited by UV light will rapidly convert into excimers which decay through fluorescence with a lifetime of $\mathcal{O}$(100~ns), and because the nature of this interaction is dependent on the presence of neighbouring pyrene monomers, there is a strong dependence on the number density of pyrene.
\par 
TPB films can be applied to materials such as acrylic by thermal vacuum evaporation~\cite{Broerman:2017hf}. Due to its higher vapour pressure of \SI{6e-6}{\milli\bar}~\cite{sonnefeld1983dynamic}, a pyrene film created by thermal deposition is not stable for LAr applications where a high vacuum is required before filling. However, pyrene-doped polymers, such as polystyrene (PS), have been studied extensively~\cite{Johnson,itaya1990}.
Mobility of pyrene molecules in the polymeric matrix is very strongly constrained, which is reflected by the diffusion constant of pyrene in PS, $D$=9.4$\times$10$^{-19}$~cm$^2$/s, one of the lowest ever measured~\cite{diffusion_pps}, making manufacturing films with good PLQY and vacuum stability feasible.

\subsection{Background rejection with a slow WLS}

Pulse shape discrimination (PSD) is used in LAr detectors to separate electron-recoil (ER) backgrounds, such as the beta decay of \ar, from the nuclear-recoil (NR) WIMP signal events.  Alpha decays in LAr have a scintillation time structure that is too similar to that of the expected dark matter signal for PSD to discriminate them.  While alpha decays inside the fiducial volume can be discriminated against WIMP recoils based on their much higher energy, those that occur in a region of the detector with incomplete light collection, such as in detector inlets, can be reconstruct as having low energies, so that they appear in the WIMP search region both in energy and in PSD space.
This is a limiting background in the DEAP-3600 experiment~\cite{Ajaj:2019wi}. 

The DEAP-3600 acrylic vessel (AV) has a neck at the top through which evaporated argon gas flows up toward a cooling coil to be liquefied again. This inlet contains so called "flowguides" - a set of acrylic components meant to direct the LAr and gaseous argon (GAr) flow. The flowguides are showered with LAr droplets from the cooling coil, which is believed to create a thin LAr film. Radon progeny on the surface of the flowguides undergo $\alpha$-decay, causing VUV scintillatation light in this LAr film.  Most of the  scintillation light is absorbed by the bare acrylic of the flowguides, but a small fraction of the light makes its way into the AV, where it is wavelength shifted by a layer of TPB and then detected by the inner photo-detectors. Since most of the event's scintillation light is shadowed by the flowguides, the reconstructed energy of these events is much smaller than the true energy and some of these events appear in the low-energy WIMP search region. Simulations of this configuration reproduce the unique features observed in the detector data~\cite[Section VII-D3]{Ajaj:2019wi}.

\begin{figure}[h!]
    \centering
    \includegraphics[width=0.8\linewidth]{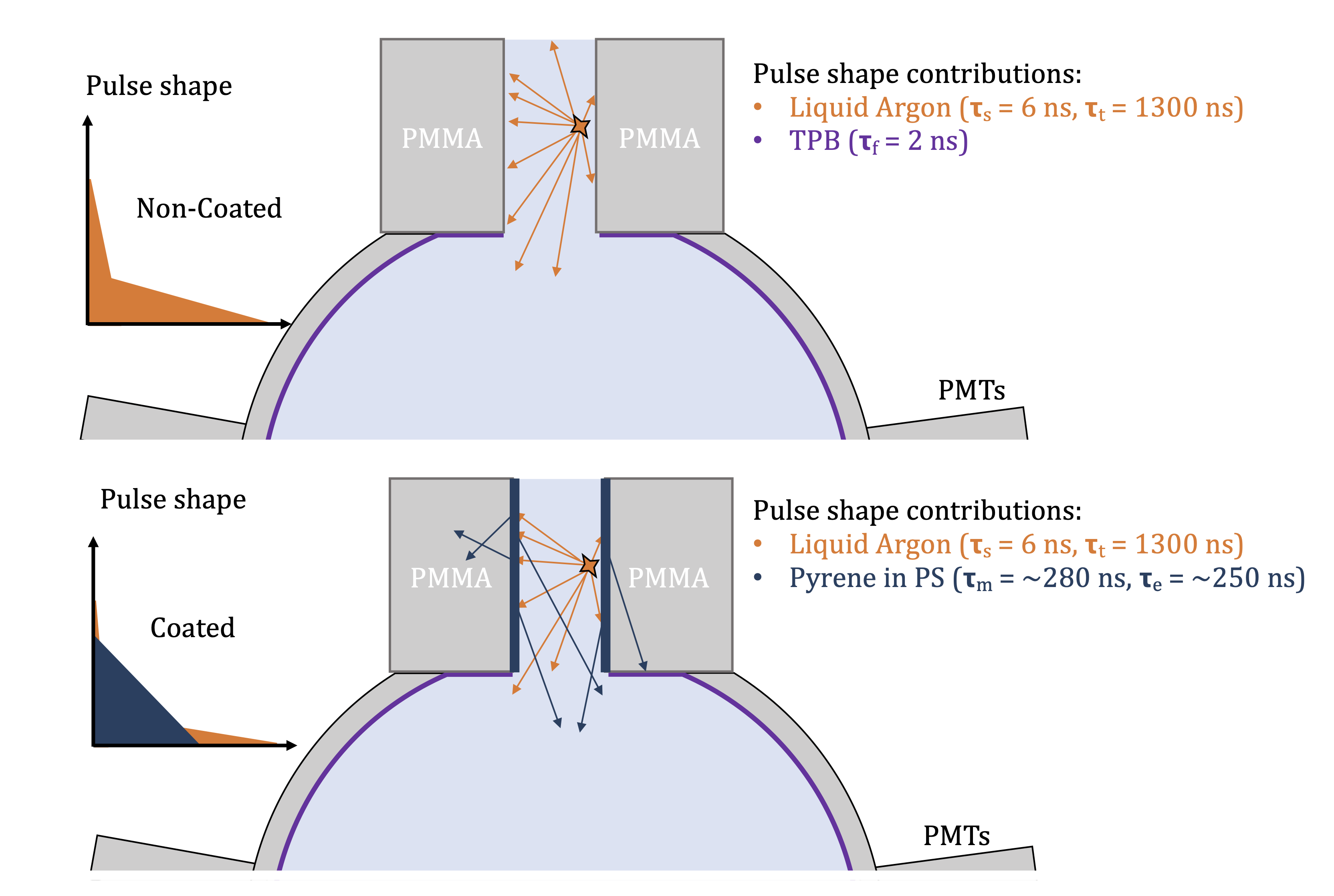}
    \caption{Top: $\alpha$ scintillation event occurs in the neck of the DEAP-3600 detector, VUV scintillation light is absorbed by the acrylic of the neck and produces a shadowed low energy, highly prompt, event that mimics a potential dark matter signal. Bottom: VUV light from $\alpha$ scintillation in argon is absorbed by the pyrene + PS film coating and is shifted to visible and delayed by the time constant of the film. This produces events with a strong "intermediate" component that can be tagged efficiently using PSD, as illustrated by the inlaid pulse shapes.}
    \label{fig:coated_Neck}
\end{figure}

The shadowed $\alpha$ scintillation events and expected pulse shape are illustrated in \autoref{fig:coated_Neck} (top). Due to the fast re-emission time of the TPB coating on the inside of the AV, the pulse shape is determined by the LAr singlet (approximately \SI{6}{\nano\second}) and triplet (approximately \SIrange{1300}{1600}{\nano\second}) excimer states.
By coating the flowguide surfaces with a thin WLS film, VUV light from $\alpha$ scintillation in LAr from this region will be wavelength shifted, so that it can be detected by the PMTs. This increases the reconstructed event energy, moving the events out of the WIMP energy region of interest. Furthermore, choosing a WLS with a slow time response shifts some prompt LAr scintillation photons to later times. This moves the events away from the WIMP search region of interest in PSD space. With its $\mathcal{O}$(100~ns) time constant, pyrene is a suitable candidate to achieve the desired distortion in the pulse shape. We will discuss the PSD power against shadowed $\alpha$'s achievable by a pyrene coat in Section~\ref{chap:results}, after describing the production and characterization of pyrene films.

\section{Film Preparation}
\label{chap:prep}
Films were prepared with 5\% through 25\%  of pyrene dissolved in polymethyl methacrylate (PMMA), polyvinyltoluene (PVT), and polystyrene, applied to acrylic by painting with a brush, by pour and set, and by dipping. Based on the characterization of the films, which is described in Section~\ref{chap:charact}, a film composed of PS, doped with 15\% pyrene (PPS film hereafter), was selected as the optimal WLS for the DEAP-3600 neck. The film preparation technique detailed here, and the characterization results presented in Section~\ref{chap:results} therefore focus on this combination. 

\subsection{Production of pyrene-doped polystyrene films}

The procedure for preparing the PPS film is detailed below. All steps except the cleaning are carried out in a nitrogen-purged glove-box to prevent oxidation of the pyrene film. To ensure the stability of the film, it is critical to keep the pyrene and the PPS film isolated from oxygen and moisture and to reduce exposure to UV light, as this was observed to have a deteriorating effect on the quality of the coating. 
All WLS films were deposited onto a sanded and cleaned acrylic substrate. The targeted dry coating thickness was achieved by controlling the concentration and amount of paint applied. Coating thickness was estimated using by measuring the mass of the applied coating, assuming a uniform coating. 

\begin{enumerate}
    \item Sonicate polystyrene beads in ultra-pure water~(UPW) and Alconox solution then rinse with UPW to remove any surface contamination
    \item Using a clean beaker, add 15\% pyrene\footnote{Sigma-Aldrich 82648-10G (CAS\# 129-00-0, fluorescence grade)} by weight
    \item Add 85\% polystyrene beads\footnote{Scientific Polymer Cat\# 400 (CAS\# 9003-53-6)} by weight 
    \item Add toluene to beaker (changing the volume of toluene relative to the solids will affect the viscosity of the film, for our application we use 1.0~g of solids to 3.0~mL of toluene for a paint-able viscosity)
    \item Placing beaker on a hot-plate with a magnetic stir stick, heat to \SI{75}{\celsius} and stir at medium velocity for 15~minutes (solids will dissolve if left in toluene without heat or stirring after several hours)
    \item Remove beaker from hot plate and let cool for 15~minutes
    \item Using a nylon bristled brush, apply coating to sample surface 
    \item If a thicker coating is desired, several coats can be applied by repeating steps 2-7 after the previous coating has set for several hours
    \item Wait 48 hours before handling: a few hours for coatings to set, the rest for the residual solvent traces to evaporate
\end{enumerate}

While not in use, samples were stored in dark and sealed containers. 

The following samples were produced for characterization:
\begin{enumerate}
    \item Single-coat and double-coat, 98\% purity PPS samples used for cryogenic tests at Carleton University, with thicknesses of 10-50~$\mu$m
    \item Single-coat, 99\% purity (\SI{9}{\micro\meter}), 98\% purity (\SI{8}{\micro\meter}), and double-coat 98\% purity (\SI{19}{\micro\meter}) PPS samples sent to the Technical University of Munich (TUM)
    \item Single-coat, 15\% pyrene of 99.9\% purity in: PVT (\SI{73}{\micro\meter}), PMMA (\SI{65}{\micro\meter}) sent to TUM
    \item Single-coat, 15\% pyrene concentration, PPS samples with purities of: 99.9\% (\SI{50}{\micro\meter}), 99\% (\SI{12}{\micro\meter}), and 98\%(\SI{15}{\micro\meter}), sent to Queen's University
    \item Single-coat, 12\% pyrene concentration, 99.9\% purity (\SI{55}{\micro\meter}) PPS sample sent to Queen's University
    \item Single-coat, 98\% purity (\SI{16}{\micro\meter}) PPS sample sent to Institute of Low Temperature and Structure Research of the Polish Academy of Sciences (ILTSR)
\end{enumerate}

All samples listed above were prepared following the above procedure. Preliminary studies were performed with samples coated by surrounding the sample acrylic substrate with a dam and flooding the sample to reach the desired thickness, this method produced very uniform coatings. The final coating method outlined here with a nylon brush was chosen for its applicability to the coatings produced for the DEAP-3600 flowguides.

\subsection{TPB reference sample production}
TPB reference samples were prepared in a thermal vacuum deposition chamber at Carleton University, described in~\cite{TPBpaper}. The TPB was loaded into a porcelain crucible wrapped with a nichrome heating element with a Pt100 temperature sensor attached. Temperature data was sent to a Lakeshore monitor that controls the current supply of the heating element to maintain a constant temperature set at the desired deposition rate. The deposition is monitored with a quartz crystal monitor located next to the sample, 18~cm above the crucible, and read out with a custom LabVIEW program. The TPB was first cycled to evaporation at \SI{240}{\celsius} (measured at the heating element) to prevent sputtering. The acrylic substrates were prepared in the same manner as those used for the PPS samples. After sanding and cleaning, the samples were loaded into the vacuum chamber. The chamber was evacuated to $<$10$^{-4}$~mbar. Deposition proceeded at a rate of 1-2~\AA/s at \SI{240}{\celsius} for 2.5~hours, resulting in a thickness of $\sim$\SI{1}{\micro\meter}. 

\par 

All TPB samples used by TUM, Queen's and ILTSR  were prepared in the same deposition and stored in a dark container within a nitrogen-purged glove box for one week until shipment to the respective institutions. All pyrene and TPB samples were stored in sealed, dark containers during shipment and in storage before measurements at each institution. Samples were exposed to air during loading and between measurements at all institutions. 


\section{Characterization of films}
\label{chap:charact}

The PPS film are used at LAr temperature (approximately \SI{87}{\kelvin}) and have to shift LAr scintillation light into the visible regime. Thus, the cryogenic stability of the coating itself, as well as the photoluminescence timing, spectrum, and quantum yield (PLQY), at 87 K are required for a full characterization of the films. The PLQY will be measured relative to the PLQY of the TPB reference samples.

\subsection{Mechanical and cryogenic stability}
Samples with varying concentrations of pyrene were produced, those with concentrations greater than 18\% pyrene by mass displayed severe crystallization of pyrene during curing and were not pursued further. 

Two different cryogenic tests were carried out; in the first, samples of pyrene with PS, PMMA, PVT were placed inside a LAr cryostat at Carleton University~\cite{Gallacher_2020} and held for 2 months. Of the 3 classes of samples, only PS samples showed no visible cracking or coating degradation. In the second test, 1" acrylic disks coated with different thicknesses of PS were slowly cooled by lowering samples into LN$_2$ vapour space until eventually submerging. No macroscopically visible cracking was observed for any of the PPS samples, regardless of thickness or number of coats. Close inspection of samples under a compound microscope showed micro-cracking of the film at the $\mathcal{O}$(1-\SI{10}{\micro\meter}) scale. These micro-cracks were stable under moderate mechanical stress, and no material loss was observed for any samples during subsequent cryogenic cycling of samples.

\subsection{Photoluminescence properties}

Measurements of photoluminescence response at cryogenic temperature and in the VUV wavelength regime are notoriously difficult and prone to systematic uncertainties. Therefore, we use three separate setups:  
the optical cryostat at Queen's University, described in~\cite{CLARK20186} and in~\cite{Benmansour_2021} (excitation with \SI{285}{\nano\meter} photons, sample at cryogenic temperatures), the VUV setup at TUM described in~\cite{Araujo:2019jd} (excitation with \SI{128}{\nano\meter} photons, sample at room temperature), and the VUV setup at ILTSR (excitation with \SI{128}{\nano\meter} photons, sample at cryogenic temperature).

\subsubsection{Decay time constant}
\label{chap:charact_queens}

The fluorescence decay time constants of 15\% PPS film samples were studied at 300~K and at 87~K using the optical cryostat at Queen's University. The setup consists of a closed-cycle optical cryostat capable of reaching temperatures between 4~K and 300~K with a pressure <$10^{-6}$~mbar. A 285~nm UV LED is set up outside the optical cryostat window to provide a $\lesssim 10$~ns FWHM excitation pulse that interacts with the sample within the cryostat chamber. For this study, the UV LED light interacts with the pyrene coating on an acrylic substrate. Outside the cryostat exit window there is a 375--650~nm broadband filter to prevent any stray UV LED light from reaching the photodetector. 
Additional filters were used in measurements to study the separate contributions from the pyrene monomer (effectively a 400~nm shortpass) and excimer (455~nm longpass) fluorescence components to the overall pyrene fluorescence.
The fluorescent light from the sample is then detected by a Hamamatsu R6095 PMT with a super-bialkali photocathode. The setup is described in detail in~\cite{Benmansour_2021}.

At each temperature, the data set consists of \SI{45}{k} waveforms of the fluorescence pulse from the pyrene sample measured by the PMT. The data analysis of these time-resolved measurements are done using the average pulse of all the waveform pulses in the data set.
Analysis over a broader range of temperatures is available elsewhere~\cite{Benmansour_2021}.

\subsubsection{Emission Spectrum}
To characterize the WLS emission spectrum and intensity, we use the TUM VUV setup, with the following modifications from \cite{Araujo:2019jd}: the deuterium lamp was replaced with model H2D2 L15094 from Hamamatsu, and for the characterization of the final film, the sample chamber and sample holder were replaced for better control of the sample angle and optimized light collection. All measurements were done at room temperature and wavelength-resolved with an OceanInsight QE65000 spectrometer. The response of the light detection system, composed of the lens, optical fiber, and spectrometer, was calibrated with a reference light source. The setup is described in detail in ~\cite{andreas_thesis}.

The following measurement protocol was followed: The monochromator entrance and exit slits were opened to \SI{5}{\milli\meter} and a wavelength of \SI{130}{\nano\meter} was selected on the monochromator. At the chosen slit width, the wavelength resolution is approximately \SI{12}{\nano\meter}. After installing the sample in the sample holder, the sample chamber was closed and the setup was evacuated to a pressure of \SI{5e-4}{\milli\bar} or better. The deuterium lamp was then turned on and reached stable light intensity after \SI{1}{\minute}. Sample spectra were measured 10 times for \SI{20}{\second} each and then averaged.

Each measurement of a WLS sample spectrum was preceded by a measurement of the TPB reference. Spectra were corrected for dark-noise and for the wavelength-dependent response function of the photon detection system, and then integrated between \SIrange{350}{650}{\nano\meter} to obtain the emission intensity. 
We consider three sources of uncertainty: a) the uncertainty on the dark noise level, b) system instability, and c) the uncertainty on the response function of the light detection system. Assuming that the TPB reference sample did not degrade over the time-frame of these measurements, we take as the system instability the RMS over the individual measurements of the TPB emission intensity. With the original sample holder, the system instability over one week of measurements was approximately 3\%.  With the new sample holder, this decreased to 1\%. The uncertainty on the response function of the photon detection system is given by the RMS over several calibrations, and is treated as a systematic uncertainty common to all measurements.

The emission spectrum was also measured with the setup at ILTSR lab, with limited resolution due to cryogenic equipment reducing light collection efficiency. 

\par

\subsubsection{Photoluminescence Yield} 

The relative PLQY of PPS at \SI{80}{\kelvin} with respect to room temperature was measured for excitation with \SI{128}{\nano\meter} photons at ILTSR. The setup consists of a McPherson vacuum mo\-nochro\-mator and spectrophotometer supplied with a deuterium lamp (EXW Herrsching L1835), a PMT (Hamamatsu R928) and a longpass filter (Schott WG320). 

The system is also equipped with an optional cold-finger cryostat (KrioSystem 03020) permitting to cool down the sample.

Furthermore, the relative PLQY of PPS with respect to TPB was extracted from the pulse shapes recorded with the Queen's optical cryostat, albeit at a higher wavelength. The amplitude of each of the \SI{45}{k} waveforms was integrated over a three-microsecond window containing the main peak and the decay. The integrated amplitude distribution for all events was then fitted by a skew normal distribution whose mean was associated with the average amount of light. 

The relative PLQY was finally also extracted from the spectra measured with the TUM VUV setup. Each sample's PLQY with respect to a TPB reference was calculated by taking the ratio of the integrated spectra in a window from 350~nm to 650~nm.
\section{Results}
\label{chap:results}

\subsection{Photoluminescence Time Constants}

\begin{figure}[h!]
    \centering
    \includegraphics[width=0.7\columnwidth]{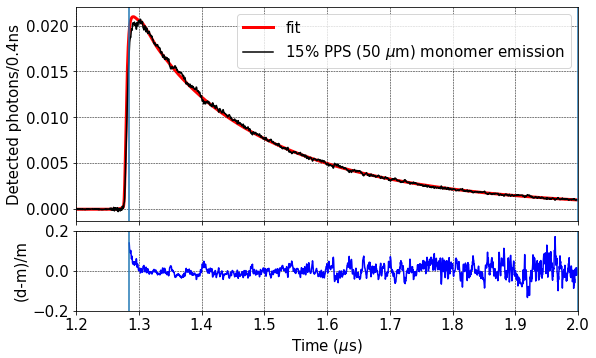}
    \caption{Fit of the pyrene monomer photoluminescence decay at 87~K using model from \autoref{eq:monomer_fit} in a [1285~ns~-~2000~ns] fitting window}
    \label{fig:monomer_fit}
\end{figure}

\begin{figure}
    \centering
    \includegraphics[width=0.7\columnwidth]{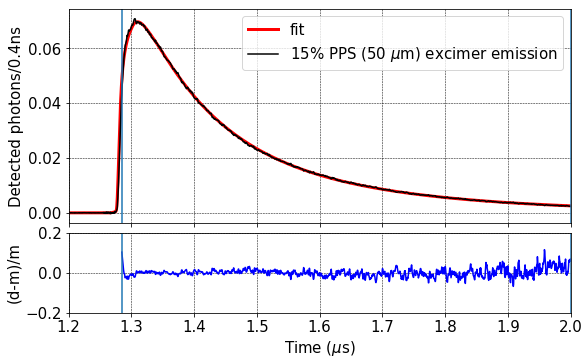}
    \caption{Fit of the pyrene excimer photoluminescence decay at 87~K using the model from \autoref{eq:excimer_fit} in a [1285~ns~-~2000~ns] fitting window}
    \label{fig:excimer_fit}
\end{figure}

The photoluminescence pulse shapes of the monomer and excimer emissions of the 15\% PPS (\SI{50}{\micro\meter}) film are fitted with a function that is the convolution of the UV excitation pulse timing with the photoluminescence decay model $i(t)$ to extract the decay time constants.
The monomer pulse shape can be represented using a decay model described in detail in~\cite{Johnson}:
\begin{equation}
\label{eq:monomer_fit}
i_m(t) = \frac{N_1}{\tau_1} \exp\Big({-\frac{t}{\tau_1}-2q\sqrt{\frac{t}{\tau_1}}}\Big),
\end{equation} 
where $q$ parametrizes the non-exponential nature of the decay of excited states caused by enhanced energy transfer at short distances~\cite{bennett_mechanisms_1968}. 
An example of the monomer decay fit at 87~K is shown in \autoref{fig:monomer_fit}.
The figure also shows the relative residuals $\frac{data-model}{model}$.
The pulse shape of the excimer-only photoluminescence is not well-described by a single exponential decay term. We use the following ad-hoc model:
\begin{equation}
\begin{split}
\label{eq:excimer_fit}
i_e(t) & = - \frac{N_{rise}}{\tau_{rise}} e^{-\frac{t}{\tau_{rise}}} + \frac{N_2}{\tau_2} e^{-\frac{t}{\tau_2}} + \frac{N_3}{\tau_3} e^{-\frac{t}{\tau_3}}
\end{split}
\end{equation}
%
%
An example of the excimer decay fit at 87~K is shown in \autoref{fig:excimer_fit}. 
Decay times obtained from those fits are summarized in \autoref{tab:queensTC}.

\begin{table}[h]
    \caption{Photoluminescence decay times and fractions of PLQY ($F_i$) contributed by the monomer and the two excimer decays for 15\%~PPS~(\SI{50}{\micro\meter})}
    \centering
    \begin{tabular}{p{0.13\linewidth}|p{0.12\linewidth}|p{0.12\linewidth}|p{0.12\linewidth} |p{0.12\linewidth} |p{0.12\linewidth}| p{0.12\linewidth}  }
    \hline
      & \multicolumn{2}{c|}{Monomer} & \multicolumn{2}{c|}{Excimer (Short)} & \multicolumn{2}{c}{Excimer (long)} \\
    \hline
    \textbf{Temp. (K)}& \textbf{$\tau_1$ (ns)} & \textbf{$F_1$ (\%)}   &\textbf{$\tau_2$ (ns)}  &\textbf{$F_2$  (\%)}  & \textbf{$\tau_3$ (ns)} & \textbf{$F_3$ (\%)} \\
    \hline
      300    & \num[separate-uncertainty = false]{214 \pm 11} & \num[separate-uncertainty = false]{23 \pm 6} &  \num[separate-uncertainty = false]{91 \pm 5} & \num[separate-uncertainty = false]{36 \pm 9} & \num[separate-uncertainty = false]{194 \pm 10}&  \num[separate-uncertainty = false]{43 \pm 10} \\
      87   & \num[separate-uncertainty = false]{279 \pm 14} & \num[separate-uncertainty = false]{52 \pm 13} &\num[separate-uncertainty = false]{105 \pm 5} & \num[separate-uncertainty = false]{36 \pm 9} & \num[separate-uncertainty = false]{249 \pm 12}& \num[separate-uncertainty = false]{31 \pm 8} \\
          \hline
    \end{tabular}
    \label{tab:queensTC}
\end{table}
Errors of $\pm 5\%$  were attributed to the decay time values to account for the systematic and statistical errors from the fit.

The combined efficiencies of the optical filters of the PMT in the monomer and excimer wavelength ranges are $\epsilon_m$ and $\epsilon_e$, respectively. At each temperature, we calculate the total photoluminescence intensity as $I_t = N_1/\epsilon_m + (N_2 + N_3)/\epsilon_e$. The fractional contribution of the monomer and the two excimer states to the PLQY is then calculated as 
    $F_1 = \frac{N_1/\epsilon_m}{I_t}$,
    $F_2 = \frac{N_2/\epsilon_e}{I_t}$,
    $F_3 = \frac{N_3/\epsilon_e}{I_t}$
and summarized in \autoref{tab:queensTC}. The contribution of the rise-time ($N_{rise}$) is treated as a systematic uncertainty.



\subsection{Photoluminescence Yield and Emission Spectrum}

\begin{figure}
    \centering
    \includegraphics[width=0.7\columnwidth]{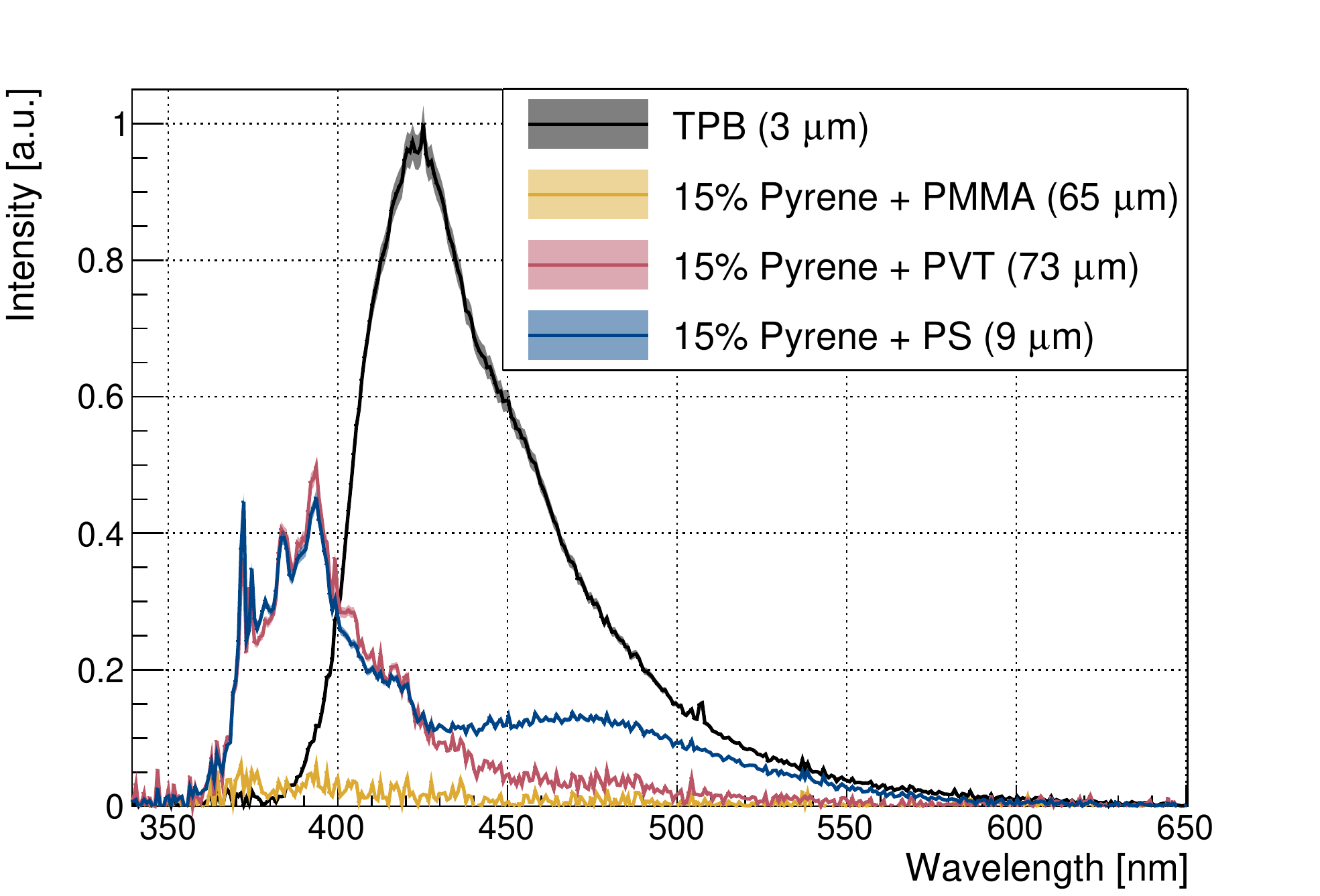}
    \caption{The spectra recorded at room temperature for three samples of 15\% pyrene in different polymeric matrices with \SI{128}{\nano\meter} excitation are shown together with the TPB reference spectrum. The shaded regions indicate the size of the statistical uncertainties related to the dark noise correction and the system fluctuation.}
    \label{fig:wlsspectra}
\end{figure}

The room temperature spectra of three WLS samples and of the TPB reference excited with \SI{128}{\nano\meter} photons using the TUM setup are shown in \autoref{fig:wlsspectra}. The spectral shapes and intensities of the three PPS samples studied at TUM matched within uncertainties. We show results from the sample with \SI{9}{\micro\meter} thickness here.

Each sample's relative PLQY with respect to a TPB reference was calculated by taking the ratio of the integrated spectra in a window from 350~nm to 650~nm. The relative yields for a set of samples using different polymeric matrix materials are shown in \autoref{tab:all}. 

\begin{table}[h]
    \caption{This table summarizes the measurements of the relative photoluminescence yield of 15\% pyrene in PMMA, PVT, and PS (the latter is referred to as `PPS' in the text) samples with respect to the TPB reference sample.Values in grey are taken from samples as measured at TUM, due to limited resolution at ILTSR, with scaling applied as measured.}
    \centering
    \begin{tabular}{p{0.28\linewidth}|p{0.1\linewidth}|p{0.1\linewidth}|p{0.14\linewidth}|p{0.10\linewidth}}
        \hline

      \textbf{Sample}   & \multicolumn{4}{c}{\textbf{ PLQY w.r.t. TPB [\%]}} \\
        &\multicolumn{2}{c}{ \SIrange{80}{87}{\kelvin}}& \multicolumn{2}{c}{ room temp.}\\
        &  \SI{128}{nm} &  \SI{285}{nm} & \SI{128}{nm} & \SI{285}{nm} \\
        \hline 
       Pyrene + PMMA (\SI{65}{\micro\meter})  & & & 3.7(0.5) & \\
       Pyrene + PVT (\SI{73}{\micro\meter})  & & & 34.9(2.7) &\\
       \hline
       PPS (\SI{9}{\micro\meter})  & & & 46.4(2.9) & \\
       PPS (\SI{50}{\micro\meter}) & & 53(6) & & 59(6)  \\
       PPS (\SI{16}{\micro\meter})& 36(6) & & \textcolor{gray}{\num{46.4}} & \\
        \hline
       \end{tabular}
    \label{tab:all}
\end{table}

\begin{table}[h]
    \caption{This table summarizes the intensity ratio of the excimer and monomer emission at \SI{128}{nm} and \ledwavelength{}, for room and LAr temperatures. Values in grey are taken from samples as measured at TUM, due to limited resolution at ILTSR.}
    \centering
    \begin{tabular}{p{0.25\linewidth}|p{0.12\linewidth}|p{0.12\linewidth}|p{0.15\linewidth}|p{0.11\linewidth}}
    \hline
      \textbf{Sample}   & \multicolumn{4}{c}{\textbf{ Excimer/Monomer Intensity}} \\
     &\multicolumn{2}{c}{ \SIrange{80}{87}{\kelvin}}& \multicolumn{2}{c}{ room temp.}\\
        &  \SI{128}{nm} &  \SI{285}{nm} & \SI{128}{nm} & \SI{285}{nm} \\
        \hline       
       PPS (\SI{9}{\micro\meter})  & & &  \SI{0.43}(0.05) &  \\
       PPS (\SI{50}{\micro\meter}) & & \SI{0.9}(0.3) & & \SI{3.4}(1.1) \\
       PPS (\SI{16}{\micro\meter})& \num[separate-uncertainty = false]{0.47} & & \textcolor{gray}{\num{0.43}} & \\
       \hline
       \end{tabular}
    \label{tab:ratio}
\end{table}

We use the pulse shapes measured in the Queen's optical cryostat to study the evolution of the PLQY with temperatures. The pulse shape from each of the \SI{45}{k} recorded events was integrated in a three-microsecond window containing the main peak and the decay. The integrated intensity distribution for all events was then fitted by a skew normal distribution whose mean was associated with the average amount of light.

For PPS (\SI{50}{\micro\meter}), the PLQY increases by \num[separate-uncertainty = false]{7 \pm 1}\% when cooling from room temperature to 87~K, while the TPB film's PLQY increases by \num[separate-uncertainty = false]{19 \pm 2}\%.

Given a PPS/TPB PLQY ratio of \num{59 \pm 6}\%  at RT, this translates into a PPS/TPB PLQY ratio of \num{53 \pm 6}\% at \SI{87}{\kelvin}. This result is shown in the second-to-last column of \autoref{tab:all}.

\begin{figure}[ht]
    \centering
    \includegraphics[width=0.7\linewidth]{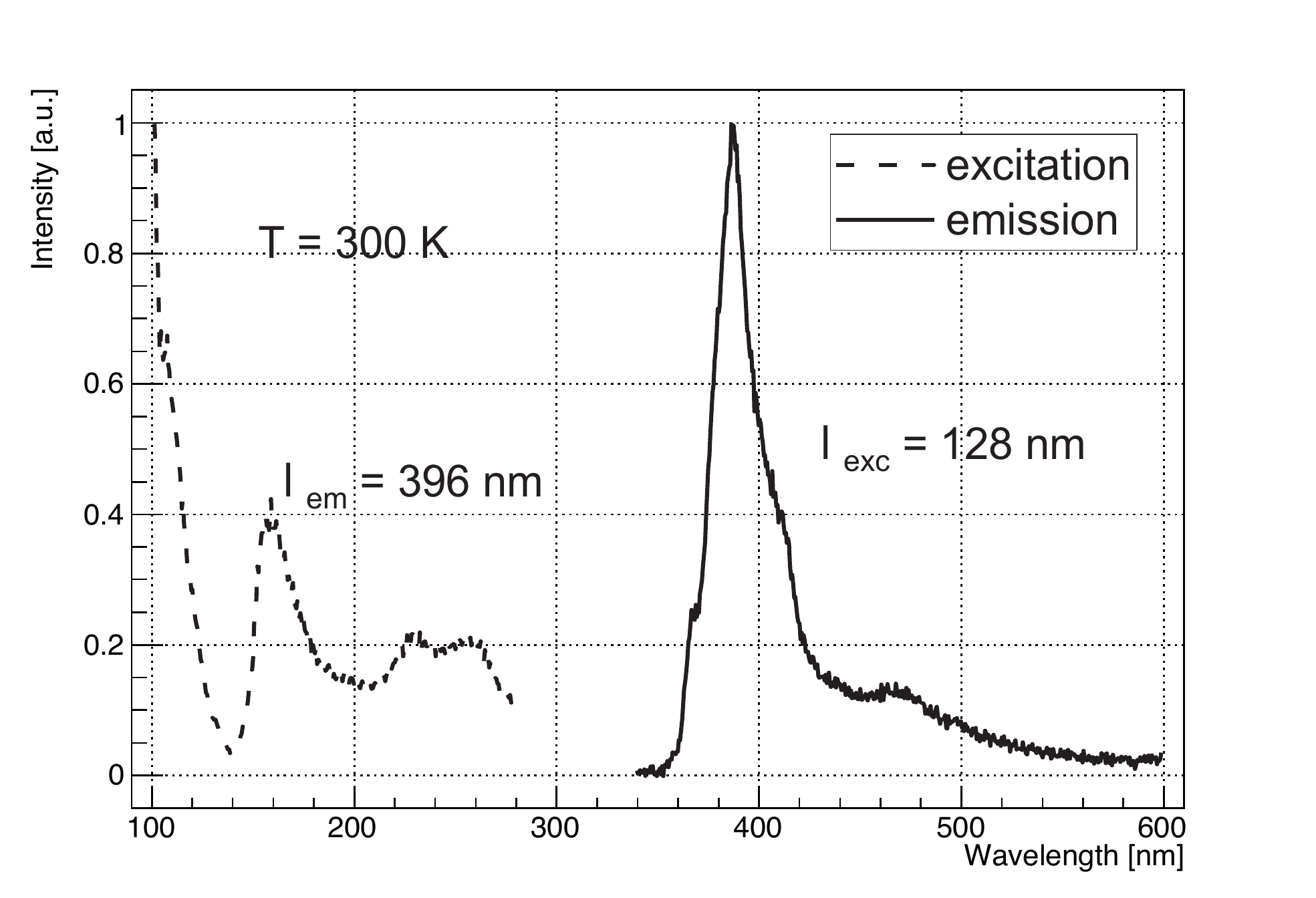}
    \caption{Excitation and emission spectra for a 15\% PPS coating, measured at room temperature. The excitation spectrum (left) shows the monomer emission intensity at 396~nm as a function of the excitation wavelength. The emission spectrum (right) was measured at 128~nm.}
    \label{fig:exc-emm}
\end{figure}

Using the ILTSR setup, the monomer excitation spectrum shown in \autoref{fig:exc-emm} exhibits strong increase below 150~nm as well as less pronounced bands at 175, 245 and 270~nm. Consistent with other measurements, the emission spectrum recorded at 128~nm consists of dominant monomer and excimer peaks at 396 and 475~nm, respectively.

\begin{figure}[ht]
    \centering
    \includegraphics[width=0.7\linewidth]{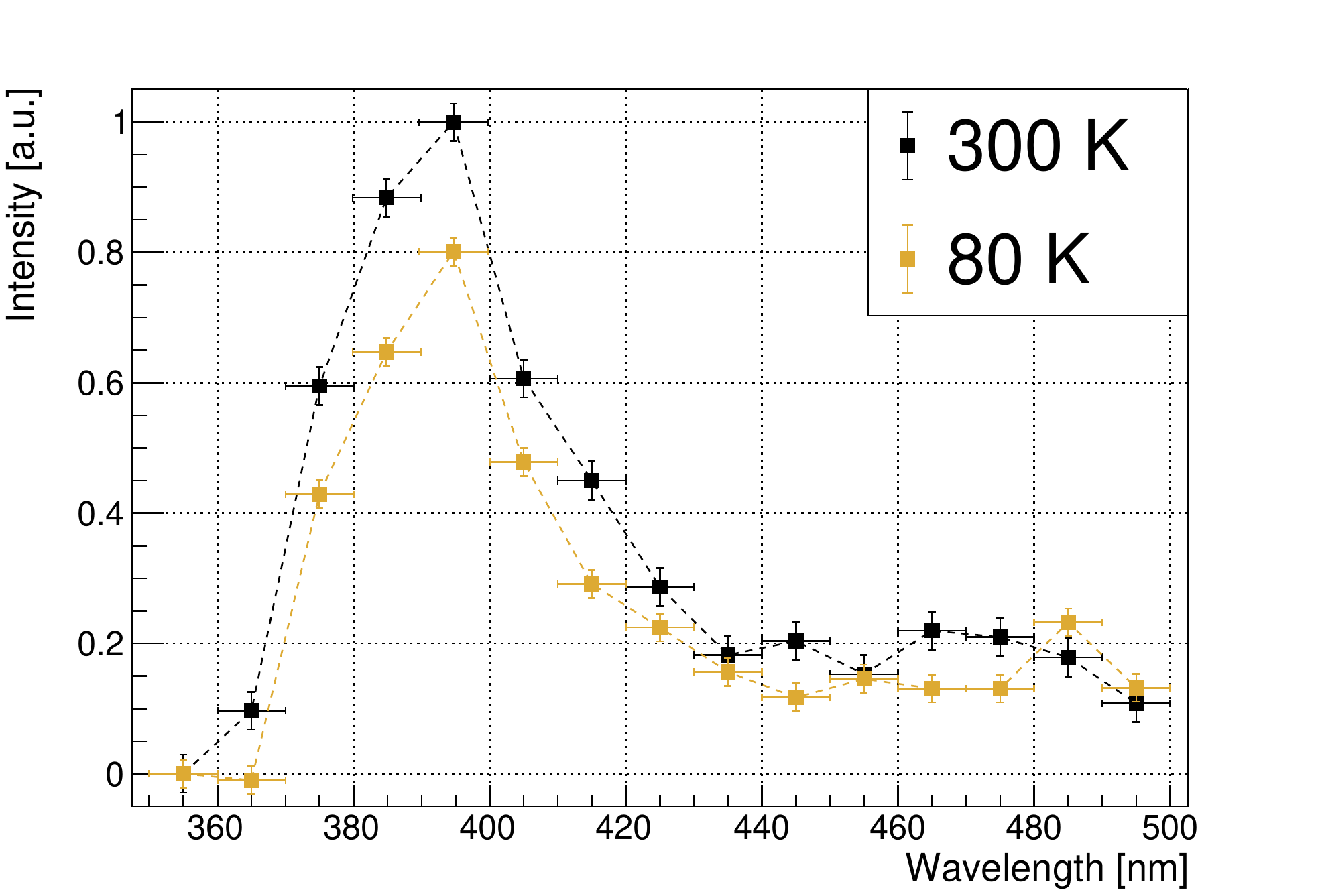}
    \caption{Emission spectra of a 15\% PPS (\SI{16}{\micro\meter}) sample at 128~nm excitation wavelength, at 300~K and 80~K, measured at ILTSR. Collected with the sample installed in the cryostat.}
    \label{fig:cryo128nm}
\end{figure}

The PPS spectra measured at RT and at \SI{80}{\kelvin} using the ILTSR setup are shown in \autoref{fig:exc-emm} and \autoref{fig:cryo128nm}. The low resolution of the low temperature spectra is due to the sample cooling system, which degrades the light collection efficiency. The temperature dependence of the response of the setup was calibrated using samples of TPB and polyethylene naphthalate (PEN) WLS's, which have known~\cite{Francini:2013noa,2pac} ratios of low-temperature and room-temperature PLQY at 128~nm excitation. Consistent calibration factors were obtained from both materials.

The relative PLQY change between room temperature and 80~K at 128~nm excitation wavelength  was obtained by integrating the spectra in the 370--420~nm range for the monomer, 450--500~nm for the excimer emission, and 370--500~nm for the overall factor. 

The PLQY of the monomer at \SI{80}{\kelvin} decreased by \num{-25 \pm 10}\%, while that of the excimer decreased by \num{-17\pm14}\%, making the intensity of the total emission spectrum decrease by \num{-23 \pm 11}\% compared to the RT value.

\subsection{Predicting the background discrimination efficiency} \label{sec:toymc}
A toy Monte-Carlo simulation was carried out to estimate the expected background discrimination efficiency with a PPS coated inlet, as illustrated in \autoref{fig:coated_Neck}. Input parameters for the simulation are summarized in \autoref{tab:tmc}.

\begin{table*}[h!]
    \caption{Input parameters for simulation}
    \centering
    \begin{tabular}{p{0.2\linewidth}|p{0.25\linewidth}|p{0.2\linewidth}|p{0.2\linewidth}}
      \hline
      \textbf{Input Component}   & \textbf{Model} &\textbf{Parameters} & \textbf{Reference} \\
      \hhline{-|-|-|-}
      LAr pulse shape   & 2-exponential + Intermediate & $\tau_s = 2.2\,ns$, $\tau_t = 1445\,ns$, $\tau_I = 75.5\,ns$  & \cite{DEAPCollaboration:2020hx} \\
      PPS pulse shape   & 3-exponential & \autoref{tab:queensTC} \& \autoref{tab:queens_contrib} & This article\\
      TPB pulse shape  & Volt-Laustriat + 2-exponential & See Table 1 in Ref. & \cite{DEAPCollaboration:2020hx}\\
      $\alpha$ Energy  & Mono-energetic & 5.3 MeV & \cite{BROWN20181}\\
      NR Energy  & Uniform Distribution & 50-250 keV & Assumed \\
      Shadowing Fraction & Uniform Distribution & 50-100$\%$ & Assumed\\
      PMT Efficiency & Measurement & Mean of 30\% & \cite{pmtpaper} \\
      TPB Emission Spectrum & Measurement &  Mode at $\sim$420 nm & This article (\autoref{fig:wlsspectra})\\ 
       PPS Emission Spectrum & Measurement &  Mode at $\sim$390 nm & This article (\autoref{fig:wlsspectra})\\ 
       PPS PLQY & Measurement &  0.46/0.36 & This article (\autoref{tab:all})\\
      \hline
    \end{tabular}
    \label{tab:tmc}
\end{table*}

\subsubsection{Simulation details}

The simulation assumes that the full energy of the 5.3 MeV $\alpha$ particle is stopped in the LAr at the inlet, which was confirmed for a \SI{50}{\micro\meter} LAr film with SRIM \cite{srim}. The energy is converted into an expected number of photons based on the liquid argon light yield of $\sim$40k~ph/MeV with an $\alpha$ quenching factor of 0.71~\cite{hitachi2019properties}. The number of photons produced for each $\alpha$ event is sampled from a Gaussian with a mean equal to the number of expected photons $\times$ quenched energy deposition and a sigma equal to the square root of the mean. Since true $\alpha$ events from the inlet will have an admixture of both LAr and wavelength shifted light from the PPS coating, the number of photons produced is split into two parts by sampling a "shadow fraction" from a uniform distribution from 50\% to 100\%. The "shadowed fraction"  represents the solid angle effect of the inlet, where the shadowing fraction represents the shadowing from the inlet geometry, and "non-shadowed" photons that represent the light from the inlet that enters the detector bulk directly without interacting with the WLS coating. All photons are given a delay time sampled from a liquid argon pulse shape described by~\cite{DEAPCollaboration:2020hx} with parameters tuned for NR-like signals. Assuming that TPB is a 100\% efficient wavelength shifter~\cite{FLOURNOY1994349}, non-shadowed photons are then given a TPB wavelength and delay time sampled from distributions as described in \autoref{tab:tmc}. Shadowed photons have a probability of being shifted by the PPS film, using the value of PLQY from this article, shown in \autoref{tab:queensTC}, photons that are successfully shifted are then given a delay time sampled from the pulse shape given in Eq.~\ref{eq:monomer_fit}, and Eq.~\ref{eq:excimer_fit} with parameters given in \autoref{tab:queensTC} and a wavelength sampled from the PPS distribution shown in \autoref{fig:wlsspectra} for PPS (\SI{9}{\micro\meter}). Shadowed photons are then given a 50\% probability of being transmitted into the detector and subsequently detected, corresponding to the solid angle effect of being in the inlet. A 400~nm cut-off is imposed on all photons to simulate the absorption from the acrylic light-guides and acrylic vessel, and geometric broadening is imposed by adding a delay time to each photon sampled from a Gaussian with parameters from~\cite{DEAPCollaboration:2020hx}. After imposing these conditions, the wavelength of each photon is used to determine the detection probability using the wavelength dependent efficiency of the Hamamatsu R5912-HQE PMTs quoted by DEAP-3600~\cite{pmtpaper}, if the photon is successfully detected its hit-time is added to a data-structure for subsequent analysis. Each detected photon has a probability to produce an after-pulse with parameters given by~\cite{DEAPCollaboration:2020hx}, if an after-pulse is registered an additional hit is triggered with a delay time sampled from the after-pulsing distribution in~\cite{DEAPCollaboration:2020hx}, modelled by a sum of Gaussian distributions. 
\par
For comparison to WIMP-like NR LAr scintillation signals for background rejection efficiency determination, the same simulation was carried out for a 50-250~keV uniform energy deposition of NR-like signals, with no shadowing or PPS photons produced. 

\subsubsection{Simulation Results}

\begin{figure}[h!]
    \centering
    \includegraphics[width=0.7\linewidth]{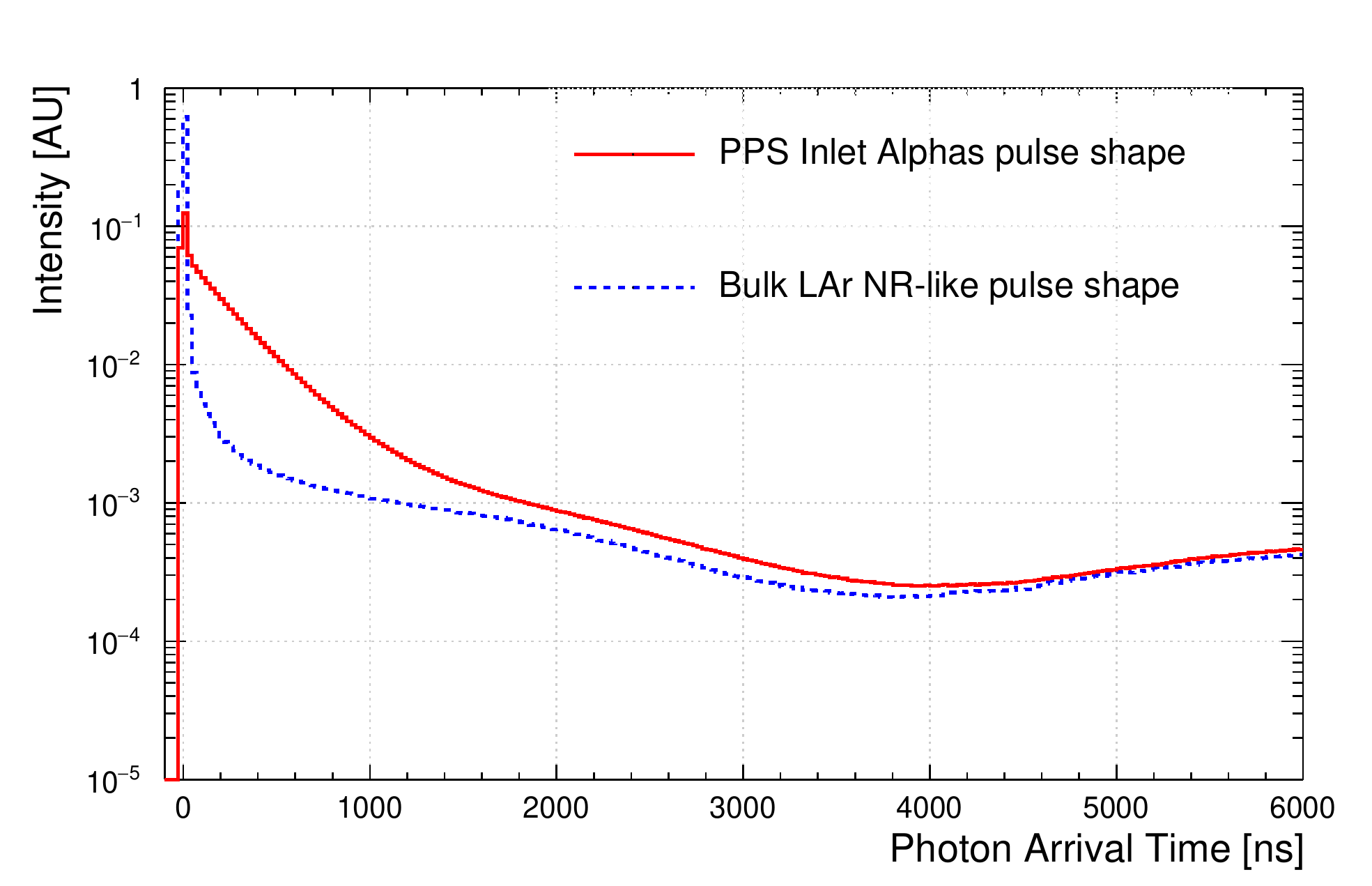}
    \caption{Comparison of pulse shapes for NR-like LAr events and inlet $\alpha$'s with a PPS coating from TMC, where both distributions are normalized to unit integral. The distributions include contributions from the LAr pulse shape, TPB re-emission and PMT effects. The bump at 2000~ns and the rise in intensity starting at \SI{4000}{\nano\second} is due to PMT afterpulsing~\cite{DEAPCollaboration:2020hx}. }
    \label{fig:tmc_ps}
\end{figure}

Comparison of the simulated pulse shapes for inlet $\alpha$'s with a PPS coating as described in this note to an expected pure LAr pulse shape for WIMP-like NR signals are shown in \autoref{fig:tmc_ps}. 

A common PSD parameter is defined as the fraction of total light detected in a ``prompt" window \autoref{eq:fprompt},
\begin{equation}
\label{eq:fprompt}
    \fp = \frac{\sum_{t_{start}}^{t_{prompt}}N(t)}{\sum_{t_{start}}^{t_{end}} N(t)}
\end{equation}
where for discrimination against \ar{} beta decays: $t_{start}$=\SI{-28}{\nano\second}, $t_{prompt}$=\SI{60}{\nano\second}, and $t_{end}$=\SI{10000}{\nano\second} (for full details, refer to~\cite{DEAPCollaboration:2020hx}). For discrimination against shadowed $\alpha$'s with a PPS coating on the flowguides, we use $t_{prompt}$=\SI{40}{\nano\second} and call this discriminator `Fpyrene'.

The Fpyrene parameter windows were tuned by scanning over integration windows for both NR-like and PPS inlet-$\alpha$ simulations, while to first-order window boundaries can be found graphically by looking for the crossing points of the normalized PDF's of the pulse shapes in \autoref{fig:tmc_ps}. 

\begin{figure}[ht]
    \centering
    \includegraphics[width=0.7\linewidth]{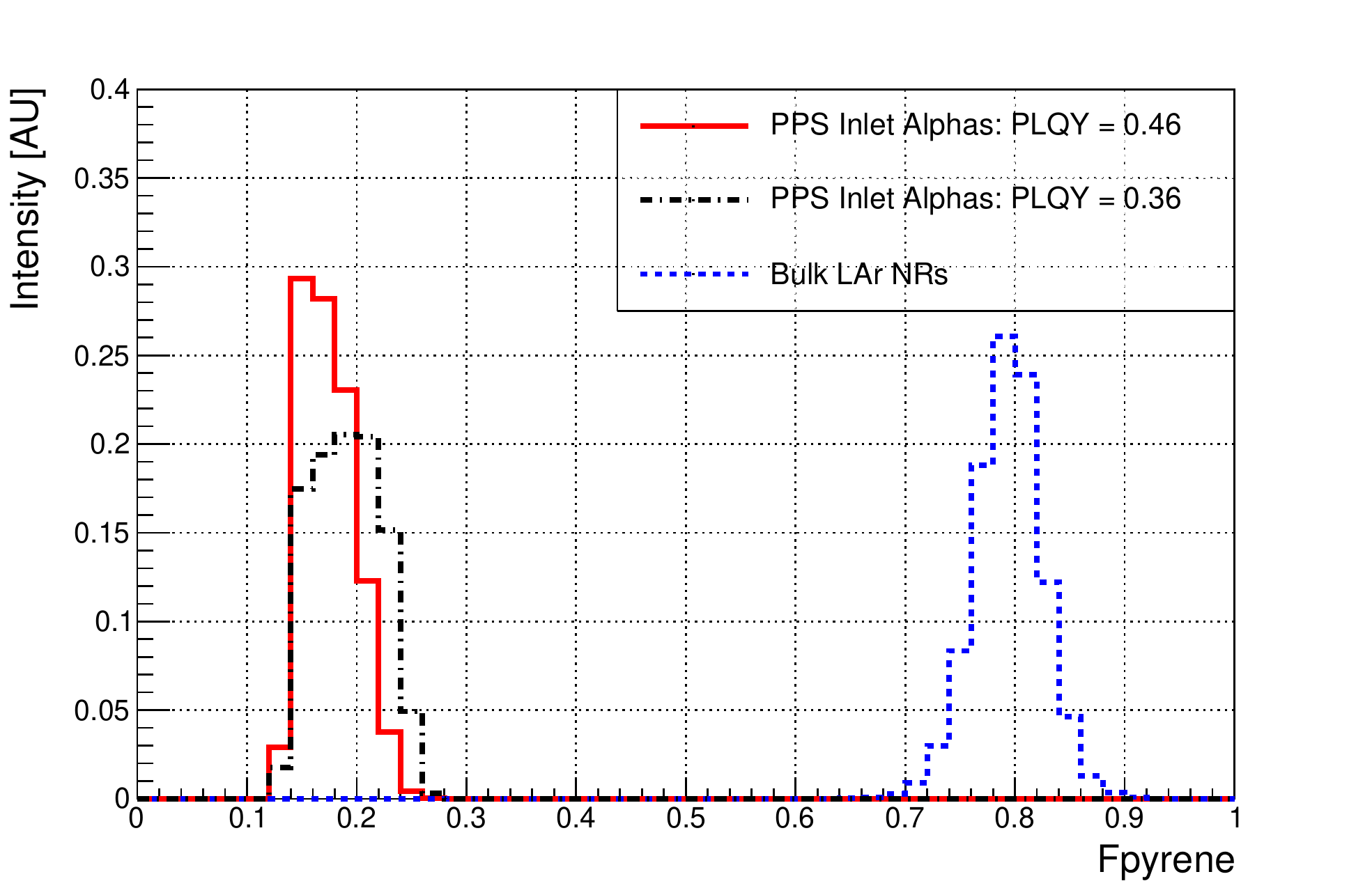}
    \caption{Fpyrene distributions for pure LAr NR-like signals and low-energy shadowed PPS coated inlet $\alpha$'s. Both nominal PLQY and lower bound for inlet $\alpha$'s are shown, with lower bound from ILTSR, all distributions are normalized to unit integral}
    \label{fig:tmc_fpyrene}
\end{figure}

The distributions of Fpyrene for simulated NR-like signals and PPS inlet-$\alpha$'s are shown in~\autoref{fig:tmc_fpyrene}. To avoid assumptions about the accuracy of the energy scale of the toy simulation, we only consider the PPS inlet-$\alpha$'s that fall in the lowest quartile of the PPS inlet-$\alpha$ energy distribution (measured in number of hit photons), which is nearest to the region of interest for WIMP searches. As there is an inverse relationship between the reconstructed energy of an event and the amount of shadowing present for that event, which with a PPS-coated inlet leads to an increase in "pyrene-like" slow photons in the PMT hit distribution, the lowest energy events reconstruct lower in Fpyrene.  

The PSD leakage fraction is defined as the fraction of PPS inlet $\alpha$'s that will leak into the region of NR-like LAr events, and is described in \autoref{eq:tmc_discrim}. We consider the full energy distribution of NR-like LAr events, and assume 100\% acceptance.

\begin{equation}
\label{eq:tmc_discrim}
    LF_{90\% C.L} =  \frac{\Big(\sum_{i = \epsilon}^{N_{bins}} F_{pyrene}^{PPS}[i]\Big)_{90\% \, C.L}}{N_{Evts}}
\end{equation}

The upper limit on leakage into the ${}^{40}$Ar NR region of interest for nominal PLQY is 
\boldmath
\leakage, assuming 100\% NR signal acceptance. The conservative lower bound on the PPS PLQY with reduction measured at ILTSR was applied and is shown in \autoref{fig:tmc_fpyrene}, no events are observed to leak into the region of interest. 
\unboldmath

\section{Discussion}
\label{chap:discuss}

We produced wavelength-shifting coatings that consist of pyrene embedded in a polymer matrix (PMMA, PVT, or PS) on a sanded acrylic substrate. The coatings were tested for their wavelength-shifting response and for mechanical stability in a cryogenic environment, and a coating composed of 15$\%$ pyrene and 85$\%$ polystyrene (PPS) by weight was identified as a candidate for use in the DEAP-3600 detector upgrade. 

After undergoing cryogenic cycling, the candidate coating did show some microscopic crazing at the $\mathcal{O}$(\SI{1}{\micro\meter}) scale. The crazing did not lead to de-lamination or material loss even when subjected to further cryogenic cycling, and the optical transmission at visible wavelengths was unaffected. 
The mechanical stability of the PPS films after cryogenic cycling indicates that the bond between the PPS film and the sanded acrylic substrate is stronger than the PPS itself. This is likely caused by the use of toluene as the solvent for the PPS film, which will also dissolve the acrylic surface and would lead to a material interface between the PPS film and acrylic that is a homogenous mixture of the PPS and acrylic. This acts as a strong bond for the material.  Therefore, we expect this coating to be stable under operation in a LAr environment. 

Three different setups were used to characterize the photoluminescence response of the PPS film, because of their complementary capabilities and as cross-check for each other's results. For technical reasons, PPS films of different thicknesses were characterized in the three setups. We checked that the coating thickness has negligible influence on the results. The wavelength ranges that could be used for the PPS monomer and excimer emission are also slightly different between setups. This introduces a discrepancy up to 6\% in the reported relative PLQY values.

We find that the PPS film has a pulse shape characterized by the exponential decays of a monomer and of two excimer states. The relative abundances and lifetimes of the three states are presented in \autoref{tab:queensTC}. At \SI{87}{\kelvin} and for excitation with \ledwavelength{} photons, the monomer, which has a lifetime of \SI{279}{\nano\second}, dominates the emission intensity, followed by the excimer state with \SI{249}{\nano\second} lifetime. Thus, the majority of photons are re-emitted with a suitable timing approximately 50~times larger than the LAr singlet excimer decay time.  The decay times of the pyrene excited states are independent of the excitation wavelength, as long as this wavelength is below the emission spectrum of the pyrene. However, the fractional contribution of the excimer and monomer states to the total pulse-shape as outlined in \autoref{tab:queensTC} can be different at the LAr scintillation wavelength of \SI{128}{nm} if there is a pyrene density gradient in the film, because the \ledwavelength{} photons penetrate deeper into the film than the \SI{128}{nm} photons. We observe that the excimer over monomer intensity ratio decreases strongly when going from \ledwavelength{} to \SI{128}{nm} excitation light, both for the measurements at room temperature and at cryogenic temperature (since only the Queen's setup can separate the two excimer components, only their sum is considered for comparisons with the other results). This indicates that the density of pyrene molecules is lower near the film surface, so that when the pyrene there is excited, the excitation has less chance of finding a dimer to form an excimer with before it decays.

We measured the PLQY of the PPS film relative to the PLQY of a TPB film that is similar to the film on the DEAP-3600 detector vessel. The PLQY of TPB has previously been reported to increase by \num{22 \pm 13}$\%$ when going from RT to \SI{87}{\kelvin}~\cite{Francini:2013noa} under excitation with \SI{128}{\nano\meter} photons. This is consistent with the \num{19 \pm 2}\% increase measured using the Queen's optical cyrostat. The photon detection efficiency of the ILTSR device has a temperature dependence that had to be calibrated by determining a scaling factor that requires the TPB emission intensity ratio between RT and \SI{80}{\kelvin} to match the literature value. The calibration was confirmed by measuring the PLQY of a second WLS with known behaviour at cryogenic temperatures. The discrepancy between the behaviour of the PPS film's PLQY at cryogenic temperatures between the Queen's and the ILTSR measurement can have several explanations:
1) The behaviour of the PPS film at cryogenic temperatures could differ between the two excitation wavelengths. 
2) Sample-to-sample differences lead to a wavelength or temperature-dependent effect, for example, different scattering behaviour at the surfaces of the samples. 
3) The monomer's contribution to the total emission intensity increases at low temperatures, especially under excitation with \ledwavelength{} light. The monomer's wavelength range covers a region where both the PMT used at Queen's and the spectrometer used at ILTSR have steep efficiency curves. A small inaccuracy in the calibration of the response of the light detection system can thus lead to large differences in results.

We consider the difference between the two measured values for the relative PLQY at cryogenic temperature as the size of the systematic uncertainty on this value. 

The PPS and TPB emission spectra measured here and shown in \autoref{fig:wlsspectra} are used as input to a toy Monte Carlo simulation. The RT spectra for \SI{128}{\nano\meter} excitation photons are used because the change of the spectral shape for this excitation wavelength is not significant when going to cryogenic temperatures, and the resolution is better than for the spectra shown in \autoref{fig:cryo128nm}. The simulation also accounts for the measured PPS pulse shape and the PPS PLQY relative to TPB, as well as for the wavelength-dependent photon detection efficiency and instrumental noise of the DEAP-3600 light detection system. The simulations indicate that by using pulse shape discrimination together with the PPS film on the detector inlet, the inlet-$\alpha$ background rate can be reduced by a factor of better than \num{1.2e-5}, while not rejecting any potential WIMP events. 

The effect of the newly introduced inlet coatings on events in the inner detector can be estimated by considering the relative surface area of the pyrene coated surface to the TPB coated inner vessel.  The neck of the DEAP acrylic vessel subtends a solid angle of $\sim 1\%$ from the centre of the vessel and since to first order only half of the WLS light from the PPS coating will transit back down into the detector, the expected contribution of ``pyrene-like'' photons to an inner detector event is minimal. If we assume activities as reported in Ref.~\cite{Ajaj:2019wi}, without evaluating second order effects and cut efficiencies, this technique is sufficient to operate the DEAP-3600 detector for $\sim$4 years without the inlet-$\alpha$ background leaking into a DM signal region.

\section{Conclusion}
\label{chap:conc}

We presented a method that can be used to achieve the background goals in the DEAP-3600 detector with regard to so-called `inlet-$\alpha$s'. The detector inlet is a region of incomplete light collection, and some $\alpha$-decay events in this region reconstruct in the region of interest for WIMP search. By coating the surface of the inlet with a WLS that has a re-emission time much longer than the LAr singlet state, and then using pulse shape discrimination, we show that it is possible to shift the position in both energy and PSD space where these background events reconstruct, so that they are no longer in the WIMP search window.

The candidate coating for use in this method is composed of 15$\%$ pyrene and 85$\%$ polystyrene (PPS) by weight, and we describe a procedure for producing a cryogenically stable PPS coating. Characterization results of the photoluminescence response of PPS films were used as input into a toy Monte Carlo simulation and confirm that the photoluminescence quantum yield, emission spectrum, and emission timing are suitable to suppressing inlet-$\alpha$'s.

Future studies will characterize the reflectivity, transparency and refractive index of the film under incident visible light for input in the full-scale GEANT4/ RAT~\cite{G4,RAT} based MC used by DEAP-3600 to determine the inlet-$\alpha$ suppression factor with better precision.

\subsection*{Acknowledgements}
We thank the Natural Sciences and Engineering Research Council of Canada, the Canadian Foundation
for Innovation (CFI), the Ontario Ministry of Research
and Innovation (MRI), Queen’s University, Carleton University, the Canada First Research Excellence Fund, and the Arthur B. McDonald Canadian Astroparticle Research Institute. We acknowledge support from  the International Research Agenda Programme AstroCeNT (MAB\allowbreak/2018\allowbreak/7) funded by the Foundation for Polish Science (FNP) from the European Regional Development Fund. AstroCeNT and Technical University of Munich (TUM) cooperation is supported from the EU’s Horizon 2020 research and innovation program under grant agreement No~962480 (DarkWave).
\appendix

\printcredits

\bibliographystyle{elsarticle-num}

\bibliography{refs}
\end{document}